\documentclass[10pt,journal,letterpaper,compsoc]{IEEEtran}
\usepackage{times}
\usepackage[final]{graphicx}
\usepackage[reqno]{amsmath}
\usepackage{amsfonts}

\usepackage{times,amsmath,epsfig}
\usepackage{latexsym,amssymb}
\usepackage{cite}

\usepackage{graphicx}         
\usepackage{color}            
\usepackage{epic}             
\usepackage{eepic}            
\usepackage{rotating}         
\usepackage{type1cm}          
\usepackage{epsfig}

\newcommand{\beq}{\begin{equation}}
\newcommand{\enq}{\end{equation}}
\newcommand{\beqa}{\begin{eqnarray}}
\newcommand{\enqa}{\end{eqnarray}}
\newcommand{\beqn}{\begin{eqnarray*}}
\newcommand{\enqn}{\end{eqnarray*}}

\newcommand{\qed}{\hfill $\Box$}

\begin{document}

\title{Inter-Sensing Time Optimization in Cognitive Radio Networks}

\author{Omar~Mehanna~
        and Ahmed~Sultan~
\IEEEcompsocitemizethanks{
\IEEEcompsocthanksitem O. Mehanna is with the Department
of Electrical and Computer Engineering, University of Minnesota, Minneapolis, USA.\protect\\
E-mail: {meha0006@umn.edu}
\IEEEcompsocthanksitem A. Sultan is with the Wireless Intelligent Networks Center (WINC),
Nile University, Cairo, Egypt.\protect\\
E-mail: {asultan@nileuniversity.edu.eg}
\IEEEcompsocthanksitem This work was partially supported by the Egyptian NTRA. 
}
\thanks{}}

\markboth{Submitted to IEEE Transactions on Mobile Computing}%
{Shell \MakeLowercase{\textit{et al.}}: Bare Demo of IEEEtran.cls for Journals}

\IEEEcompsoctitleabstractindextext{%
\begin{abstract}
We consider a set of primary channels that operate in an unslotted fashion, switching activity at random times. A secondary user senses the primary channels searching for transmission opportunities. If a channel is sensed to be free, the secondary terminal transmits, and if sensed to be busy, the secondary transmitter remains silent. We solve the problem of determining the optimal time after which a primary channel needs to be sensed again depending on the sensing outcome. The objective is to find the inter-sensing times such that the mean secondary throughput is maximized while imposing a constraint over the maximum tolerable interference inflicted on the primary network. Our numerical results show that by optimizing the sensing-dependent inter-sensing times, our proposed scheme reduces the impact of sensing errors caused by false alarm and misdetection and outperforms the case of a single sensing period.

\end{abstract}

}

\maketitle


\section{Introduction}
The radio spectrum resource is of fundamental importance
to wireless communication. Recent reports show that most
available spectrum has been allocated. However, most of licensed spectrum resources are under-utilized. This
observation has encouraged the emergence of dynamic and
opportunistic spectrum access concepts, where secondary (unlicensed)
users (SU) equipped with cognitive radios are allowed to
opportunistically access the spectrum as long as they do not
interfere with primary (licensed) users (PU). To achieve this goal, the
secondary users must monitor the primary traffic in order to
identify spectrum holes or opportunities which can be exploited to
transfer data \cite{Haykin}.

There are two main
scenarios for the primary-secondary coexistence. The first is
the overlay scenario where the secondary transmitter checks
for primary activity before transmitting. The secondary user
utilizes a certain resource, such as a frequency channel, only
when it is unused by the primary network. The second scenario
is the underlay system where simultaneous transmission
is allowed to occur so long as the interference caused by
secondary transmission on the primary receiving terminals
is limited below a certain level determined by the required
primary quality of service. For both scenarios, the cognitive MAC protocol should
continuously make efficient decisions on which channels to sense and
access in order to obtain the most benefit from the available
spectrum opportunities. Previous work on the design of cognitive MAC protocols has considered two distinct scenarios. In the first, the primary network is slotted (e.g., \cite{HangSu}, \cite{Capacity}, \cite{Zhao}, \cite{ZhaoWhitle}, \cite{Lifeng}  \cite{Motamedi} and \cite{Blind}) whereas a continuous structure (un-slotted) of the primary channels is adopted in the second set of works (e.g., \cite{HyoilKim}, \cite{HyoilKim2}, \cite{OptimalSensing}, \cite{newZhao} and \cite{Ding}).

In this work, we focus on the un-slotted primary network set-up. We consider different sensing and transmission capabilities for the SU. Using the Orthogonal Frequency Division Multiplexing (OFDM) technique  with adaptive and selective allocation of OFDM subcarriers, any subset of licensed channels can be utilized at the same time. We first assume that the SU radio can transmit on any combination of the primary  channels at the same time, whereas it is capable of sensing only one channel at any instance. Although sensing may be less demanding than transmission, we use this assumption to compare our results with those presented in \cite{HyoilKim}. The SU aims at maximizing its throughput (\emph{i.e.,} maximizing the opportunities discovered and accessed in all primary channels) while imposing minimal interference to the primary network. Specifically, after the SU senses the primary channels, it is required to find the optimal duration until the next sensing time. During this duration, secondary transmission takes place on the channels that are sensed to be free while channels that are sensed busy are left idle till the next sensing event. A similar model is adopted in \cite{HyoilKim}, where the authors have developed an optimal sensing period for each of the primary channels by optimizing the tradeoff between the sensing overhead resulting from frequent sensing of the channels and the missed opportunities in the primary channels due to infrequent sensing. However, it is assumed that if a primary transmission is resumed on a channel, the SU discovers this return, via the help of a Genie, and immediately evacuates the channel, thereby causing no interference to the primary transmissions. In this work, we relax this Genie-aided assumption and impose an interference/outage constraint on each primary channel. More importantly, we show that by optimizing different sensing durations corresponding to different possibilities of the sensing outcome of the channels, the performance can be substantially improved.

We next consider the case where the SU can sense more than one channel at the same time. Following the same objectives as the previous case, we propose a simple, less complex, myopic scheme followed  by an optimal scheme. Based on the channels sensing outcome, the optimal duration till the next sensing time is found. Finally, we consider the scenario when the SU radio can be tuned to only one channel. The SU in this case tries to access a primary channel so long as it is free. When this channel switches to busy, the SU searches the other primary channels until a free channel is identified. A similar model is adopted in \cite{HyoilKim2}, where an optimal sequence of primary channels to be sensed is proposed. This optimal sequence aimed at minimizing the average delay in finding a free channel. Here, we extend this work by finding the period a free channel shall be accessed in order to satisfy an interference/outage constraint on the primary network, which was not considered in \cite{HyoilKim2}.

%
The rest of the paper is organized as follows. Section~\ref{model}
presents our modeling assumptions. Section \ref{Formulation} provides the formulation for our problem. The proposed protocol for the limited sensing, full capabilities and limited channel access scenarios are developed in Sections \ref{limited_sensing}, \ref{full} and \ref{limited_access} respectively. Numerical results for our proposed strategies are reported in Section~\ref{numerical}. Finally, Section~\ref{conclusion} summarizes our conclusions.

\section{Network Model}\label{model}

\subsection{Primary Network}
We consider a primary network consisting of $N_P$ independent channels. The presence or absence of
primary users in each channel can be modeled as alternating time intervals of busy and free states
with random durations. For channel $i \in 1,2,\ldots , N_P$, we model the sojourn time of a busy period as a random variable $T_i^0$ with the continuous probability density function (p.d.f.) $f_{T_i^0}(y), y \geq 0$. Similarly, the p.d.f. of the sojourn time in a free period is given as $f_{T_i^1}(x), x \geq 0$. Busy and free periods are assumed to be independent and identically distributed (i.i.d.). We also assume that busy and free periods are independent of each other. Without loss of generality, we focus on exponentially distributed busy/free periods for each channel as an illustrative example:
\begin{eqnarray}
  f_{T_i^1}(x) &=& \lambda_{T_i^1} e^{- \lambda_{T_i^1} x} \\
  f_{T_i^0}(y) &=& \lambda_{T_i^0} e^{- \lambda_{T_i^0} y}
\end{eqnarray}
where $\frac{1}{\lambda_{T_i^1}}$ is the mean of the free period and $\frac{1}{\lambda_{T_i^0}}$ is the mean of the busy period. The channel utilization $u_i$ in this case is given by:
\begin{eqnarray}
u_i &=& \frac{E[T_i^0]}{ E[T_i^0] + E[T_i^1]}  \\
&=& \frac{\lambda_{T_i^1}}{(\lambda_{T_i^1}+\lambda_{T_i^0})}
\end{eqnarray}


\subsection{Secondary Pair}
The SU takes two actions: sensing and transmission. It
uses a spectrum sensor (e.g., based on energy or feature) to
determine whether the PU is idle or busy at a given time.
Of the $N_P$ primary channels, the SU can access $N_A \leq N_P$ channels simultaneously and can sense $N_S \leq N_A$ channels simultaneously. In this paper we consider the three cases of $N_S=N_A=N_P$, $N_A=N_P , N_S=1$ and $N_S=N_A=1$. We assume that the SU is equipped with a single antenna that can be used for either sensing or transmission. That is, the secondary transmitter does not transmit while sensing any channel. The actual channel $i$'s state $S_i(t)$ at time $t$ is:
\begin{equation*}
    S_i(t) = \begin{cases} 0 & \text{if channel is busy} \\
    1 & \text{if channel is free} \end{cases}
\end{equation*}
whereas due to errors in the sensing, channel $i$'s state at time $t$ according to the sensing outcome at the secondary transmitter is:
\begin{equation*}
    \bar{S_i}(t) = \begin{cases} 0 & \text{if channel is sensed busy} \\
    1 & \text{if channel is sensed free} \end{cases}
\end{equation*}
The sensing vector $\Omega(t) = [\bar{S_1}(t), \ldots, \bar{S}_{N_S}(t)]$ captures the sensing outcomes of the $N_S$ sensed channels. If the secondary transmitter decides that a free channel is busy, it refrains from transmitting, and a data transmission opportunity is lost. This is the false alarm situation,
which is characterized by the probability of false alarm $P^{FA}_i = P(\bar{S_i}=0|S_i = 1)$. On
the other hand, if the detector fails to classify a busy channel as
busy, a miss detection occurs, possibly resulting in interference with
the primary user. The probability of misdetection is denoted by
$P^{MD}_i=P(\bar{S_i}=1|S_i = 0)$. If energy detection is used as a sensing method \cite{SensingTime}, the minimum required sensing time $T_s$ that satisfies a certain desired $P^{FA}$ and $P^{MD}$ is given by:
\begin{equation}\label{T_s}
    T_s = \frac{2}{f_s} \left[Q^{-1}(P^{FA}) - Q^{-1}(1-P^{MD}) \sqrt{1 + 2 \sigma}\right]^2 \sigma^{-2}
\end{equation}
\noindent where, $f_s$ is the sampling frequency, $Q(x)$ is the tail probability of a zero-mean unit-variance Gaussian random variable and $\sigma$ is the PU signal-to-noise ratio \cite{SensingTime}. The sensing time $T_s$ is assumed to be much smaller than $E(T_i^1)$ and $E(T_i^0)$. This assumption guarantees that the primary is highly unlikely to change state during the sensing period.

We also assume that the SU knows the PU free and busy time distributions. A maximum likelihood (ML) estimation for the distributions is proposed in \cite{HyoilKim} while a Bayesian estimation method is discussed in \cite{HyoilKim2}. Gernerally speaking, learning the channel statistics may require an initial learning phase at which each channel is sensed and secondary transmission is disallowed. There are also learning algorithms that allow tracking of slowly-varying channel parameters without the need for periodic learning phases that waste some of the available secondary transmission opportunities.


\section{Problem Formulation}\label{Formulation}

Our goal is to find the optimal access strategy that maximizes the throughput for SU while satisfying the PU intereference/outage constraints for each channel. In other words, after the SU senses the primary channels, it is required to find the optimal duration until the next sensing time. During this duration, secondary transmission takes place on the channels that are sensed to be free while channels that are sensed busy are left un-accessed till the next sensing event.

Since the SU depends only on sensing a channel at specific times to identify the channel's state, it cannot track the exact state transition of each channel. Hence, the free portion of time between the actual state transition from busy to free until the SU discovers this transition cannot be utilized. In addition, some free periods may remain undiscovered at all if sensing is infrequent. Following \cite{HyoilKim}, the unexplored opportunities are quantified as the average fraction of time during which channel $i$'s vacancy is not discovered by the SU . On the other hand, the transition of primary activity from free to busy on a channel utilized by the SU causes interference to the primary and secondary receivers until the SU realizes this transition. This interference is quantified as the average fraction of time at which channel $i$ is used simultaneously by the primary and secondary terminals. We assume here that concurrent transmission inevitably leads to packet loss. If we take into consideration the channels between the primary and secondary transmitters and receivers, this may not be the case. If, for instance, the channel between the secondary transmitter and receiver is in deep fade, the transmitted packet would be lost even if the primary is completely silent during the whole transmission period. Alternatively, if the channel between the primary transmitter and secondary receiver is weak, then the secondary packets may survive collisions with primary transmissions due to the small interference level. In this work, we do not account for channel gains.

Note that blindly increasing the sensing frequency to reduce interference and discover more opportunities is not desirable because the SU must suspend the use of the discovered free channel(s) in order to sense other channels. This is due to the assumption that data transmission and sensing cannot take place at the same time with one antenna. The sensing overhead is defined as the average fraction of time during which channel $i$'s discovered opportunities are interrupted due to the need for channel sensing \cite{HyoilKim}. This trade-off will be captured in the construction of our objective function which is used to find the optimal transmission/no-access duration between any two sensing events.

Before delving into your optimization problem, we need first to find expressions for $\delta_i^1(t)$, the expected time in which a channel is free during the time between $t_s$ and $t_s + t$ provided that $S_i(t_s) = 1$, and $\delta_i^0(t)$, the expected time in which a channel is free during the time between $t_s$ and $t_s + t$ provided that $S_i(t_s) = 0$. Based on the theory of alternating renewal processes, if channel $i$ is free at time $t_s$, the remaining time $\tilde{x}$ for the channel to be in the same free state can be shown to have the p.d.f. $\frac{1-F_{T_i^{1}}(x)}{E[T_i^{1}]}$, where $F_{T_i^{1}}(x)$ is the c.d.f. of the free period \cite{Cox}. Similarly, if channel $i$ is busy at $t_s$, the remaining time $\tilde{y}$ for the channel to be in the same busy state has the p.d.f. $\frac{1-F_{T_i^{0}}(y)}{E[T_i^{0}]}$. Therefore, it can be easily shown that:
\begin{eqnarray}
  \nonumber \delta_i^1(t) &=& t \int_t^\infty \frac{1-F_{T_i^1}(x)}{E[T_i^1]} dx\\
  \label{delta1} & & +  \int_0^t \frac{1-F_{T_i^1}(x)}{E[T_i^1]} (x+\tilde{\delta}_i^0(t-x)) dx\\
  \label{delta0} \delta_i^0(t) &=&  \int_0^t \frac{1-F_{T_i^0}(y)}{E[T_i^0]} \tilde{\delta}_i^1(t-y) dy
\end{eqnarray}

\noindent where $\tilde{\delta}_i^1(t)$ and $\tilde{\delta}_i^0(t)$ are the same as $\delta_i^1(t)$ and $\delta_i^0(t)$ if the change in state happens exactly at $t_s$. That is,
\begin{eqnarray}
  \nonumber \tilde{\delta}_i^1(t) &=& t \int_t^\infty f_{T_i^1}(x) dx\\
  \label{delta_1} & & +  \int_0^t  f_{T_i^1}(x) (x+\tilde{\delta}_i^0(t-x)) dx\\
  \label{delta_0} \tilde{\delta}_i^0(t) &=& \int_0^t f_{T_i^0}(y) \tilde{\delta}_i^1(t-y) dy
\end{eqnarray}

Using Laplace transform, $\delta_i^1(t)$ and $\delta_i^0(t)$ for exponentially distributed busy/free periods can be obtained as: (see Appendix A for a complete derivation)
\begin{eqnarray}
   \delta^0_i(t)  & = & (1-u_i) \cdot \left(t+ \frac{ e^{-(\lambda_{T_i^0}+\lambda_{T_i^1})t} - 1}{ (\lambda_{T_i^0}+\lambda_{T_i^1}) }\right) \\
   \delta^1_i(t) & = &   t-u_i \cdot \left(t+ \frac{ e^{-(\lambda_{T_i^0}+\lambda_{T_i^1})t} - 1}{ (\lambda_{T_i^0}+ \lambda_{T_i^1}) }\right)
\end{eqnarray}

Therefore, if $\bar{S_i}(t_s)=S_i(t_s)=0$, channel $i$ is un-accessed by the SU for some duration $t$, and the unexplored opportunities during the time between $t_s$ and $t_s + t$ is $ T_i^U(t) =  \delta_i^0(t)$,  while if $\bar{S_i}(t_s)=0$ but $S_i(t_s)=1$ we have: $ T_i^U(t) =  \delta_i^1(t)$. Similarly, if $\bar{S_i}(t_s)=S_i(t_s)=1$, the SU will transmit on channel $i$ for some duration $t$, and the interference during the time between $t_s$ and $t_s + t$ is given by: $T_i^I(t) =  t-\delta_i^1(t)$, while if $\bar{S_i}(t_s)= 1$ but $S_i(t_s)=1$ we have: $T_i^I(t) =  t-\delta_i^0(t)$

An important result that will be used in the formulation of our optimization problem is the probability that
a primary channel is free at time $t_s+ t$ where its state $S(t_s)$ at time $t_s$ is known i.e., $P_i^{11}(t) = P(S_i(t_s + t) = 1 | S_i(t_s)=1)$ and $P^{01}(t) = P(S_i(t_s + t) = 1 | S_i(t_s)=0)$. According to the renewal theory, we only need the most
recent sample from each channel. An expression for arbitrary distributions of busy and free periods is given in  \cite{Cox} and \cite{HyoilKim}. For exponentially distributed busy/free periods:
\begin{eqnarray}
  P^{11}_i(t) &=& (1-u_i) + u_i e^{- ( \lambda_{T_i^1} + \lambda_{T_i^0}) t} \label{P11}\\
  P^{01}_i(t) &=& (1-u_i)-(1-u_i)e^{-(\lambda_{T_i^1}+\lambda_{T_i^0})t } \label{P01}
\end{eqnarray}

\section{limited sensing ($N_S=1$ , $N_A=N_P$)}\label{limited_sensing}
In this section we assume that although the SU is capable of transmitting on all $N_P$ channels at the same time, it is capable of sensing only one channel at a time. Though sensing may be less demanding than transmission and, hence, it is not obvious why the secondary would sense one channel at a time while it can send on all, we consider this case to compare our results with those presented in \cite{HyoilKim}. We also consider that sensing errors take place during secondary operation.

In order to sense any channel, the transmission taking place on any other channels is paused till the end of the sensing event. The proposed algorithm relies on the novel idea of using two sensing periods for each channel: free sensing period $T_i^{F}$ if $\bar{S}_i(t) = 1$, during which secondary transmission takes place on channel $i$, and busy sensing period $T_i^{B}$ if $\bar{S}_i(t) = 0$, during which no secondary activity takes place on channel $i$. Therefore, our optimization task is to identify the optimal sensing periods $T_i^{F*}$, $T_i^{B*}$ for each channel, that maximize the total throughput for the SU on the $N$ channels while satisfying the PU interference constraint on each channel.

Assuming independent channels, we model each channel as a four-state Markov chain. The four states are defined as follows:\\ {\bf State 1}: the channel is busy and is correctly sensed to be busy, {\it i.e.}, ${S}_i(t) = 0$ and $\bar{S}_i(t) = 0$.\\
{\bf State 2} : the channel is busy but a miss detection occurs and it is sensed as free. Here we have ${S}_i(t) = 0$ and $\bar{S}_i(t) = 1$.\\ {\bf State 3} : ${S}_i(t) = 1$ and $\bar{S}_i(t) = 0$ which means that a false alarm occurs and a free channel is sensed as busy.\\ {\bf State 4}: the channel is sensed to be free and it is actually free, {\it i.e.}, ${S}_i(t) = 1$ and  $\bar{S}_i(t) = 1$.\\

Note that we embed the sensing outcome into the definition of channel state. The four-state Markov probability transition matrix of each channel is given by (\ref{M}).
\begin{figure*}
\begin{equation} \label{M}
M_i=
\left[
  \begin{array}{cccc}
    (1-P^{01}_i(T_i^{B}))(1-P_i^{MD}) & (1-P^{01}_i(T_i^{B}))P_i^{MD} & P^{01}_i(T_i^{B})P_i^{FA} & P^{01}_i(T_i^{B})(1-P_i^{FA}) \\
    (1-P^{01}_i(T_i^{F}))(1-P_i^{MD}) & (1-P^{01}_i(T_i^{F}))P_i^{MD} & P^{01}_i(T_i^{F})P_i^{FA} & P^{01}_i(T_i^{F})(1-P_i^{FA}) \\
    (1-P^{11}_i(T_i^{B}))(1-P_i^{MD}) & (1-P^{11}_i(T_i^{B}))P_i^{MD} & P^{11}_i(T_i^{B})P_i^{FA} & P^{11}_i(T_i^{B})(1-P_i^{FA}) \\
    (1-P^{11}_i(T_i^{F}))(1-P_i^{MD}) & (1-P^{11}_i(T_i^{F}))P_i^{MD} & P^{11}_i(T_i^{F})P_i^{FA} & P^{11}_i(T_i^{F})(1-P_i^{FA})
  \end{array}
\right]
\end{equation}
\end{figure*}
\noindent Element $(m,n)$ in this transition matrix denotes the probability of making a transition from State $m$ to State $n$. For instance, element $(1,2)$ is the product of two terms. The first is the probability of the actual channel state remaining busy. Since $P^{01}$ denotes that probability of making the transition from busy to free, the probability of remaining at the busy state is given by $(1-P^{01}_i(T_i^{B}))$. Note that since the initial state is State 1 for which the channel is sensed to be busy, the argument of function $P^{01}$ is $T_i^{B}$. The second term of element $(1,2)$ is $P_i^{MD}$ as State 2 denotes the occurrence of miss detection. We assume that the channel transition from state to state is independent of the sensing process and its outcome.

The steady state of the four-state Markov chain of channel $i$, denoted by $\Pi_i = [\pi_i(0,0),\pi_i(0,1),\pi_i(1,0),\pi_i(1,1)]$, can be found by solving $\Pi_i \, M =\Pi_i$ with $\pi_i(0,0)+\pi_i(0,1)+\pi_i(1,0)+\pi_i(1,1)=1$. Probability $\pi_i(0,0)$, for example, is the steady state probability of the channel $i$ being busy and sensed as such, {\it i.e.}, $S_i= 0$ and $\bar{S}_i = 0$.



The secondary terminal suspends transmission when it senses any primary channel. This interrupts the transmission process and lowers the throughput. The sensing overhead is defined as the expected fraction of time in which the un-interfered secondary transmission on channel $i$ is interrupted by the need for sensing. Thus, when  $\bar{S}_i= S_i=1$, the sensing overhead incurred on channel $i$ is given by
\begin{equation}
\label{overhead}
T^O_i\left(T_i^F\right)=\delta^1_i \left(T_i^F\right) \sum_{j = 1}^{N_P} \frac{T_s}{T_j^{F}}
\end{equation}
\noindent When $\bar{S}_i = 1$ but $S_i= 0$, the sensing overhead has an expression similar to (\ref{overhead}), but with $\delta^1$ replaced by $\delta^0$.

The expected normalized throughput corresponds to the expected time of transmission without interference or interruption due to sensing. It is given by
\begin{eqnarray}
\nonumber R &=& \sum\limits_{i=1}^{N_P}  \frac{\pi_i(1,1)}{\mu_i}  \delta_i^1(T_i^{F}) \left(1-\sum_{j = 1}^{N_P} \frac{T_s}{T_j^{F}}\right) \\
    & & + \frac{\pi_i(0,1)}{\mu_i}  \delta_i^0(T_i^{F}) \left(1-\sum_{j = 1}^{N_P} \frac{T_s}{T_j^{F}}\right)
\end{eqnarray}
where $\mu_i$ (normalization factor) is the average time between sensing events for channel $i$ and is given by:
\begin{equation}
    \mu_i = (\pi_i(0,0) + \pi_i(1,0))T_i^{B} + (\pi_i(0,1)+\pi_i(1,1))T_i^{F}
\end{equation}


Given a maximum interference constraint per primary channel, $T_i^{I_{\rm max}} $, our optimization problem can be expressed as follows:
\begin{eqnarray*}
	& & \text{Find: } T_i^{F*} , T_i^{B*}\\
    & & \text{that maximize: } R \\
    & & \text{subject to: } T_i^I(T_i^{F},T_i^{B}) \leq T_i^{I_{\rm max}} \: ,\:  i = 1, \ldots , N
\end{eqnarray*}
\noindent where $T_i^I(T_i^{F},T_i^{B})$ is the average time during which the primary and secondary terminals concurrently transmit on channel $i$. This average interference time is given by
\begin{eqnarray}
T_i^I(T_i^{F},T_i^{B})&=&\frac{\pi_i(1,1)}{\mu_i} (T^F_i-\delta_i^1(T^F_i)) + \nonumber \\
& &\frac{\pi_i(0,1)}{\mu_i} (T^F_i-\delta_i^0(T^F_i))
\end{eqnarray}
The optimization problem is non-convex,
hence, we do exhaustive search to numerically obtain $T_i^{F*}$  and  $T_i^{B*}$ (as demonstrated by our numerical results in Section~\ref{numerical}).

\section{Full capabilities ($N_S=N_A=N_P$)}\label{full}

Here we assume that the SU can sense and transmit on any combination of the primary channels simultaneously $N_S=N_A=N_P$. Since the secondary transmitter cannot sense any channel while in transmission, the transmission taking place on any channels is stopped during the sensing event.
Now, for each of the $2^{N_A}$ possibilities of the sensing vector $\Omega$, the objective is to find the optimal duration $T_p(\Omega)$ during which the secondary transmission takes place on the channels that are sensed to be free while channels that are sensed busy are left un-accessed. At the end of $T_p(\Omega)$ all $N_P$ channels are sensed again to capture a new vector $\Omega$ and find a new optimal duration $T_p(\Omega)$. We consider two schemes below. The first is a simple myopic scheme that aims at maximizing the immediate reward, whereas the second aims at maximizing the total expected throughput. A maximum interference constraint per primary channel $T_i^{I_{\rm max}}$ is imposed in both schemes.

As defined previously, the sensing overhead is the expected fraction of time in which the un-interfered secondary transmission on channel $i$ is interrupted by the need of sensing. Thus, if $\bar{S_i}=S_i=1$, $T_i^O (t) =  \delta_i^1(t) \frac{T_s}{t}$, while if $\bar{S_i}=1$ and $S_i=0$, $T_i^O (t) =  \delta_i^0(t) \frac{T_s}{t}$. Note that in contrast with (\ref{overhead}), there is no summation as all channels are sensed simultaneously.

\subsection{Myopic Scheme}

Given any realization of the sensing vector $\Omega(t)$ after sensing the $N_P$ primary channels, the total expected normalized throughput for a transmit duration $T_p\left(\Omega\right)$ is:
\begin{equation}
   \sum\limits_{i=1}^{N_A} \bar{S_i}(t)\frac{( T_p(\Omega) - T_i^I(T_p) -  T_i^O (T_p))}{T_p(\Omega)}
\end{equation}
\noindent where the average interference time on channel $i$, $T_i^I(T_p)= (1-P_i^{FA}) \left(T_p-\delta_i^1(T_p)\right) + P_i^{MD}  (T_p-\delta_i^0(T_p))$, and  the sensing overhead $T_i^O(T_p) =   (1-P_i^{FA}) \delta_i^1(T_p) \frac{T_s}{T_p} + P_i^{MD}  \delta_i^0(T_p) \frac{T_s}{T_p}$. Since the secondary terminal refrains from transmitting if the channel is sensed to be busy and does not sense again except after time $T_p$, the time for which the primary is off during the time interval of duration $T_p$ constitutes a lost transmission opportunity. This lost or unexplored opportunity can be represented by the following term
\begin{equation}
  \sum\limits_{i=1}^{N_A} (1-\bar{S_i}(t)) \frac{T_i^U(T_p)}{T_p(\Omega)}
\end{equation}

\noindent where $T_i^U(T_p)= (1-P_i^{MD}) \delta_i^0(T_p) + P_i^{FA} \delta_i^1(T_p)$. Hence, for each of the $2^{N_A}$ possibilities of the sensing vector $\Omega$, the optimization problem is formulated as follows:
\begin{eqnarray*}
	& & \text{Find: } T_p(\Omega)\\
    & & \text{that maximize: } \sum\limits_{i=1}^{N_A} \bar{S_i}(t)\frac{( T_p(\Omega) - T_i^I(Tp) -  T_i^O (T_p))}{T_p(\Omega)} \\
    & & \qquad \qquad \qquad - (1-\bar{S_i}(t)) \frac{T_i^U(T_p)}{T_p(\Omega)} \\
    & & \text{subject to: } T_i^I(T_p) \leq T_i^{I_{\rm max}} \: ,\:  i = 1, \ldots , N_P
\end{eqnarray*}
We do exhaustive search to numerically obtain $T_p(\Omega)$ as the optimization problem is non-convex. Note that for the myopic scheme, after sensing the channels, the goal is to find out for how long to access the free channels. Since the throughput per free channel increases as the next sensing time increases, we need to consider the loss of throughput on the other busy channels (unexplored opportunities) which also increases as the next sensing time increases. Thus, the myopic scheme, in some sense, maximizes the immediate reward only.

\subsection{Optimal Scheme}
Here, we do not just maximize the per channel throughput, we also maximize the probability of sensing a channel and finding it free. This is similar to the optimization formulation in Section 4 and achieves a better performance than the myopic scheme. Taking into account all the sensing outcome possibilities for all the channels, and assuming no sensing errors for simplicity, the total expected normalized throughput can be expressed as:

\begin{eqnarray}\label{R}
  \nonumber  R &=& \sum\limits_{k=1}^{2^{N_P}} \bigg(Prob(\Omega(t) = \Omega_k) \cdot\\
     & & \sum\limits_{i=1}^{N_A} \Omega_k(i) \frac{( T_p(\Omega_k) - T_i^I(T_p) -  T_i^O (T_p))}{\mu} \bigg)
\end{eqnarray}
where $\Omega_k$ correspond to all the $2^{N_P}$ possibilities of $\Omega$, that is: $\Omega_1 = [ 0 , 0 , \ldots, 0]$, $\Omega_2 = [1 , 0 , \ldots, 0]$, $\dots$, and $\Omega_{2^{N_P}} = [ 1 ,1,\ldots, 1]$. Parameter $\mu$ is the mean time between sensing events and, in this case
\begin{equation}
\mu = \sum_{k=1}^{2^{N_P}}  Prob(\Omega(t) = \Omega_k) T_p(\Omega_k).
\end{equation}
\noindent $\Omega_k(i)$ corresponds to the $i$th element of the vector $\Omega_k$ which takes the value of either 0 or 1.

The main challenge is to find $Prob(\Omega(t) = \Omega_k)$. Next, we present the results for $N_P=2$ channels assuming no sensing errors. Let:
\begin{equation*}
    T_p = \begin{cases} T_{0,0} & \text{if }  \Omega(t) = [0 , 0] \\
    T_{0,1} & \text{if }  \Omega(t) = [0 , 1] \\
    T_{1,0} & \text{if }  \Omega(t) = [1 , 0] \\
    T_{1,1} & \text{if }  \Omega(t) = [1 , 1]  \end{cases}
\end{equation*}

\noindent The two channels form a $4$-state Markov chain with
transition probabilities obtained using (\ref{P11}) and (\ref{P01}), and the transition matrix is given by (\ref{ChMatrix}).
\begin{figure*}
    \begin{equation}\label{ChMatrix}
M=
\left[
  \begin{array}{cccc}
    (1-P^{01}_1(T_{0,0})) (1-P^{01}_2(T_{0,0}))& (1-P^{01}_1(T_{0,0})) P^{01}_2(T_{0,0}) & P^{01}_1(T_{0,0}) (1-P^{01}_2(T_{0,0})) & P^{01}_1(T_{0,0}) P^{01}_2(T_{0,0}) \\
    (1-P^{01}_1(T_{0,1})) (1-P^{11}_2(T_{0,1}))& (1-P^{01}_1(T_{0,1})) P^{11}_2(T_{0,1}) & P^{01}_1(T_{0,1}) (1-P^{11}_2(T_{0,1})) & P^{01}_1(T_{0,1}) P^{11}_2(T_{0,1}) \\
    (1-P^{11}_1(T_{1,0})) (1-P^{01}_2(T_{1,0}))& (1-P^{11}_1(T_{1,0})) P^{01}_2(T_{1,0}) & P^{11}_1(T_{1,0}) (1-P^{01}_2(T_{1,0})) & P^{11}_1(T_{1,0}) P^{01}_2(T_{1,0}) \\
    (1-P^{11}_1(T_{1,1})) (1-P^{11}_2(T_{1,1}))& (1-P^{11}_1(T_{1,1})) P^{11}_2(T_{1,1}) & P^{11}_1(T_{1,1}) (1-P^{11}_2(T_{1,1})) & P^{11}_1(T_{1,1}) P^{11}_2(T_{1,1}) \\
  \end{array}
\right]
\end{equation}
\end{figure*}

The steady state of this Markov chain $\Pi = [\pi_{0,0},\pi_{0,1},\pi_{1,0},\pi_{1,1}]$ can be found by solving $\Pi M =\Pi$ and $\pi_{0,0}+\pi_{0,1}+\pi_{1,0}+\pi_{1,1}=1$. The average time between sensing events is:
\begin{equation}
    \mu = \pi_{0,0}T_{0,0} + \pi_{0,1}T_{0,1} + \pi_{1,0}T_{1,0}+ \pi_{1,1}T_{1,1}
\end{equation}

The normalized expected throughput for the 2-channel case is given by:
\begin{eqnarray}
    \nonumber R & = & \frac{\pi_{1,0}}{\mu} \left( T_{1,0} - (T_{1,0}-\delta_1^1(T_{1,0}) ) - \frac{\delta_1^1(T_{1,0})}{T_{1,0}} T_s \right) \\
    \nonumber & & + \frac{\pi_{0,1}}{\mu}  \left( T_{0,1} - (T_{0,1}-\delta_2^1(T_{0,1}) ) -  \frac{\delta_2^1(T_{0,1})}{T_{0,1}} T_s \right)\\
    \nonumber & &  + \frac{\pi_{1,1}}{\mu}  \bigg( 2 T_{1,1}-  (T_{1,1}-\delta_1^1(T_{1,1})) -  (T_{1,1}-\delta_2^1(T_{1,1})) \\
     & & - \frac{ \delta_1^1(T_{1,1})}{T_{1,1}} T_s - \frac{ \delta_2^1(T_{1,1})}{T_{1,1}} T_s  \bigg) \\
    \nonumber & = & \frac{\pi_{1,0} \delta_1^1(T_{1,0}) }{\mu} \left(1- \frac{T_s}{T_{1,0}}\right) + \frac{\pi_{0,1} \delta_2^1(T_{0,1} ) }{\mu} \left(1- \frac{T_s}{T_{0,1}}\right) \\
     & & + \frac{\pi_{1,1}( \delta_1^1(T_{1,1}) + \delta_1^2(T_{1,1}))}{\mu} \left(1- \frac{T_s}{T_{1,1}}\right)
\end{eqnarray}


\noindent Finally, the optimization problem is formulated as follows:
\begin{eqnarray*}
	& & \text{Find: } T_{0,0}, T_{0,1}, T_{1,0}, T_{1,1}\\
    & & \text{that maximize: } R \\
    & & \text{subject to: } {T_1^I} \leq T_1^{I_{\rm max}} \: ,\:  {T_2^I} \leq T_2^{I_{\rm max}}
\end{eqnarray*}
where ${T_1^I} = \frac{\pi_{1,0}}{\mu} (T_{1,0}-\delta_1^1(T_{1,0})) + \frac{\pi_{1,1}}{\mu}(T_{1,1}-\delta_1^1(T_{1,1}))$ and ${T_2^I} = \frac{\pi_{0,1}}{\mu} (T_{0,1}-\delta_2^1(T_{0,1})) + \frac{\pi_{1,1}}{\mu}(T_{1,1}-\delta_2^1(T_{1,1}))$. Again, the problem is non-convex,
hence, we do exhaustive search to numerically obtain the optimal inter-sensing time values (as demonstrated by our numerical results in Section~\ref{numerical}).

It is noted that the size of the matrix given in (\ref{ChMatrix}) grows exponentially with the number of channels. In addition, if the sensing errors are taken into account, the number of rows and columns of are doubled.

\section{limited channel access ($N_S=N_A=1$)}\label{limited_access}
Here, we assume that the SU can transmit on only one single channel. Our goals are 1) to find the optimal transmission period upon accessing any channel in order to satisfy an interference/outage constraint on the primary network and 2) to find the optimal sequence of primary channels to be sensed to minimize the average delay in finding a free channel.

When the SU finds a channel busy, it switches between channels and senses them until a free channel is found. This is equivalent to setting $T_i^{B} = 0$ in the expressions derived in Section 4. Upon finding a vacant channel, the SU transmits for a period of $T_i^{F}$. The average interference time $T_i^I(T_i^{F})$ is given by:
\begin{eqnarray}\label{INTF2}
    \nonumber T_i^I(T_i^{F}) &=& (1-P^{FA})  \cdot \left(\frac{T_i^{F} - \delta_i^1(T_i^{F})}{T_i^{F}}\right) \\
    & &  + \quad P^{MD} \cdot \left(\frac{T_i^{F} - \delta_i^0(T_i^{F})}{T_i^{F}}\right)
\end{eqnarray}

\noindent Assuming error-free sensing, the interference $T_i^I(T_i^{F})$ for exponentially distributed busy/free periods becomes:
\begin{equation}\label{INTF_Mod}
    T_i^I(T_i^{F})=u_i \cdot \left(1+ \frac{ e^{-(\lambda_{T_i^0}+\lambda_{T_i^1})T_i^{F}} - 1}{ T_i^{F} \cdot (\lambda_{T_i^0}+\lambda_{T_i^1}) }\right)
\end{equation}

Let $T^{I_{\rm max}}$ be the maximum fraction of outage/interference that the primary users can tolerate on all primary channels. Since only one channel is accessed at a given time, satisfying the interference constraint on each single channel is sufficient for ensuring that the total interference constraint is satisfied. Hence, assuming that the SU sensed channel $i$ to be free at time $t$ (i.e., $S_i(t)=1$), the sensing period $T_i^{F}$ for each channel must satisfy the constraint: $T_i^I(T_i^{F}) \leq T^{I_{\rm max}}$. In order to maximize the SU throughput, the optimal sensing period for each channel $T_i^{F*}$ is $T_i^{F}$ which satisfies: $(T_i^I(T_i^{F}) = T^{I_{\rm max}})$.

Now, we focus on the case when a channel is sensed to be busy. We need to find the optimal sequence of channels to sense in order to find a free channel as soon as possible, thereby minimizing the average delay in finding free channels. It is shown in \cite{HyoilKim2} that in order to minimize the average delay in finding a free channel, assuming all channels have the same capacity, the SU should attempt to access the channels in descending order of the channel index $\gamma_i$, where:

\begin{equation*}
    \gamma_i = \begin{cases} \frac{P^{11}_i(t-t_s)}{T_s} & \text{if}  \quad S_i(t_s) = 1\\
    \frac{P^{01}_i(t-t_s)}{T_s} &  \text{if} \quad  S_i(t_s) = 0 \end{cases}
\end{equation*}

In brief, the proposed strategy works as follows:
\begin{enumerate}
  \item Sense the $N_P$ channels in descending order of $\gamma_i$ until a free channel is found.
  \item Access the free channel $i$ for the calculated $T_i^{F*}$.
  \item When $T_i^{F*}$ ends, recalculate $\gamma_i$ for all channels, then repeat the previous steps.
\end{enumerate}

\section{Numerical Results}\label{numerical}

\begin{figure}
  \includegraphics[width=.5 \textwidth]{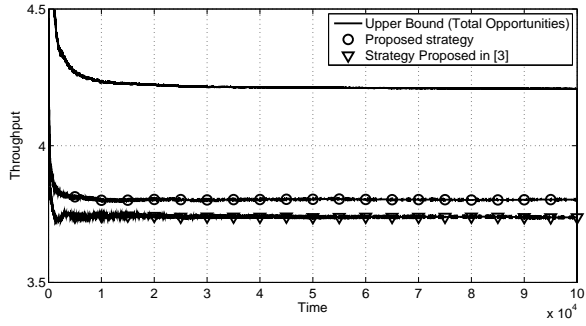}
  \caption{Performance comparison between the proposed multi-channel access strategy and the strategy proposed in \cite{HyoilKim} for an interference constraint $T_i^{I_{\rm max}} = 0.25 \, u_i$}
\label{MultiChan1}\end{figure}

\begin{figure}
  \includegraphics[width=.5 \textwidth]{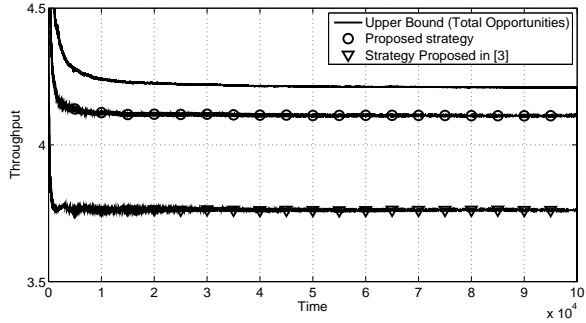}
  \caption{Performance comparison between the proposed multi-channel access strategy and the strategy proposed in \cite{HyoilKim}  for an interference constraint $T_i^{I_{\rm max}} = 0.75 \, u_i$}
\label{MultiChan2}\end{figure}

In Figures \ref{MultiChan1} and \ref{MultiChan2}, we compare the performance of the proposed strategy which adopts the sensing-dependent periods $T_i^{F}$ and $T_i^{B}$, and the strategy proposed in \cite{HyoilKim} which uses a single sensing period for each channel. In the simulations, we assume $N=5$ primary channels with exponentially distributed busy/free periods, where $\lambda_{T^1} = [0.2;0.17;0.15;0.13;0.11]$ and $\lambda_{T^0} = [1;0.9;0.8;0.7;0.6]$. Perfect sensing is assumed and the channel sensing duration is assumed to be $T_s=0.01$. The plotted results are the average over 100 simulation runs and the average throughput per time unit is plotted. The total available opportunities in the primary spectrum (upper bound on SU throughput) for the given values of $\lambda_{T^1}$ and $\lambda_{T^0}$ are: $\left(\sum\limits_{i=1}^5 (1-u_i)\right) = 4.205$ . Instead of the assumption that the SU immediately detects returning PUs and evacuate the channel, which is used in \cite{HyoilKim}, we impose an interference constraint for each primary channel.

In Figure~\ref{MultiChan1}, the interference constraint for each channel $T_i^{I_{\rm max}} = 0.25 \, u_i$. Under this assumption, our optimization method results in: $T^{F*} =[0.6133 ;   0.6800 ;   0.7637  ;  0.8714  ;  1.0148 ]$ and $T^{B*} = [ 0.3001  ;  0.3155  ;  0.3338  ;  0.3561  ;  0.3839]$. Using these values, the expected rate for the SU is given by $R=3.8068$. The optimization for the strategy proposed in \cite{HyoilKim} results in the single sensing period per channel $Tp_i^*$, where: $Tp^* = [ 0.6345  ;  0.7032  ;  0.7908  ;  0.9034  ;  1.0533]$  and an expected rate of $R=3.7531$. In Figure \ref{MultiChan2}, we set a more relaxed interference constraint for each channel $T_i^{I_{\rm max}} = 0.75 \, u_i$. Our optimization method results in: $T^{F*} =[3.8847  ;  4.3127 ;   4.8462  ;  5.5318  ;  6.4457 ]$, $T^{B*} = [ 0.2793  ;  0.2950  ;  0.3135   ; 0.3359 ;   0.3637]$, and $R=4.1085$. The optimization for the strategy proposed in \cite{HyoilKim} results in: $T_p^* = [1.0444   ; 1.1035  ;  1.1403  ;  1.1886  ;  1.2532 ]$,  and $R=3.7731$. Overall, we can see from the two figures that the throughput of the proposed strategy outperforms that of the strategy proposed in \cite{HyoilKim} at different interference constraints. However, one can observe the following trend: As the interference constraint becomes more strict (i.e., $T_i^{I_{\rm max}} \ll u_i$), more frequent sensing is required, resulting in a decreased throughput and a reduced advantage of the proposed strategy.


\begin{figure}
  \includegraphics[width=.5 \textwidth]{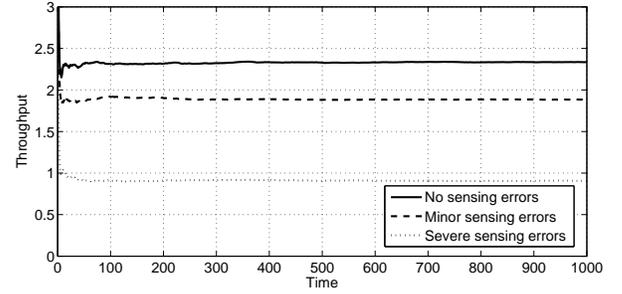}
  \caption{The effect of sensing errors on the proposed limited sensing scheme}
\label{Sensingerrors}\end{figure}

In Figure \ref{Sensingerrors}, we compare the performance of the proposed strategy which adopts the sensing-dependent periods $T_i^{F}$ and $T_i^{B}$, at different sensing error effects. In the simulations, we assume $N=3$ primary channels with exponentially distributed busy/free periods, where $\lambda_{T^1} = [0.2;0.15;0.12]/1000$ and $\lambda_{T^0} = [0.9;0.8;0.7]/1000$. The channel sensing duration is assumed to be $T_s=10$. The average throughput per time unit is plotted. The imposed interference constraint is $T_i^{I_{\rm max}} = 0.2 \, u_i$. For the case of perfect sensing (i.e., $P_i^{FA}=0$ and $P_i^{MD}=0$), our optimization method results in: $T^{F*} =[520  ; 585  ; 665]$ and $T^{B*} = [ 245 ;  285 ;  275]$. Using these values, the expected rate for the SU is given by $R=2.3228$. By introducing the effect of sensing errors, specifically $P_i^{FA}=0.2$ and $P_i^{MD}=0.1$, our optimization method results in: $T^{F*} =[285 ;  295 ;  315]$ and $T^{B*} = [ 235  ; 230  ; 235]$. Using these values, the expected rate for the SU is given by $R=1.8544$. For severe sensing errors where $P_i^{FA}=0.4$ and $P_i^{MD}=0.3$, the optimization  results in: $T^{F*} =[245  ; 245  ; 275]$ and $T^{B*} = [ 510  ; 500 ;  560]$ and the expected rate for the SU is $R=0.9377$.

\begin{figure}
  \includegraphics[width=.5 \textwidth]{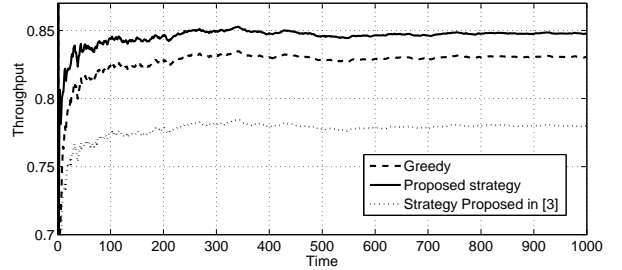}
  \caption{Performance comparison between the Full capabilities ($N_S=N_A=N_P$) proposed schemes and the scheme proposed in \cite{HyoilKim}}
\label{Comparison}\end{figure}

In Figure \ref{Comparison}, we compare the schemes proposed for the Full capabilities ($N_S=N_A=N_P$) case and the scheme proposed in \cite{HyoilKim}. In the simulations, we assume $N=2$ primary channels with exponentially distributed busy/free periods, where $\lambda_{T^1} = [0.4 ;0.7]/1000$ and $\lambda_{T^0} = [0.6 0.3]/1000$. The maximum achievable throughput (total opportunities) for these channel parameters is 0.9.  The channel sensing duration is assumed to be $T_s=10$. The average throughput per time unit is plotted. The imposed interference constraint is $T_i^{I_{\rm max}} = 0.1 \, u_i$. The proposed Myopic (Greedy) strategy optimization results in $Tp_{0,0}^* = T_s$, $Tp_{0,1}^* = 129$, $Tp_{1,0}^* = 179$ and $Tp_{1,1}^* = 204$.  Using these values, the expected rate for the SU is given by $R=0.8338$. The proposed optimal strategy optimization results in $Tp_{0,0}^* = T_s$, $Tp_{0,1}^* = 181$, $Tp_{1,0}^* = 215$ and $Tp_{1,1}^* = 650$.  Using these values, the expected rate for the SU is given by $R=0.85$. The optimization for the strategy proposed in \cite{HyoilKim} results in: $Tp^* = [301   ; 325]$,  and $R=0.783$. By relaxing the interference constraint $T_i^{I_{\rm max}} = 0.4 \, u_i$, the optimization results for the Myopic strategy and the strategy proposed in \cite{HyoilKim} do not change while the proposed optimal strategy optimization results in $Tp_{0,0}^* = T_s$, $Tp_{0,1}^* = 175$, $Tp_{1,0}^* = 241$ and $Tp_{1,1}^* = 4060$.  Using these values, the expected rate for the SU increases to $R=0.8715$.

\section{Conclusions}\label{conclusion}

We have considered in this paper an un-slotted timing structure for the primary traffic. We have investigated three scenarios that are defined according to the number of primary channels that can be sensed and accessed. In general, and in order to maximize the throughput, we have proposed an inter-sensing time scheme that depends on the sensing outcome, i.e., whether the channel is sensed to be free or busy. Our numerical results show the advantage of our proposed scheme over the case of a single sensing period in terms of maximizing the secondary throughput while imposing a constraint over the maximum tolerable interference inflicted on the primary network due to sensing errors and the random switching nature of primary activity in un-slotted systems.

Our future work would focus on the following. In this work, we have assumed that concurrent transmission automatically means that the primary and secondary packets are lost. In real-life, this needs not be the case, but depends on the physical layer channel gains. For example, if the channel between the secondary transmitter and primary receiver has a very low gain, then continuous secondary transmission would not impact the transmission taking place over the primary link. In other words, throughput-maximizing optimization problem with interference constraint can be extended to include transmission power control while incorporating the channel gains between the primary and secondary transmitters and receivers.

Another line of investigation is to find problem formulations that can be ``convexified.'' The non-convex problems are solved via computationally intensive exhaustive search, which is only possible with a small number of optimization parameters. The problem is clear for the scenario in which $N_S=N_A=N_P$ when the number of inter-sensing times is equal to $2^{N_P}$.

\appendices
\section{Derivation of $\delta_i^1(t)$ and $\delta_i^0(t)$}

By taking the Laplace transforms for equations (\ref{delta1}), (\ref{delta0}), (\ref{delta_1}) and (\ref{delta_0}):

\begin{eqnarray}
\delta^0_i(s) &=& \frac{\mathbb{F}_{T_i^0}(s) \tilde{\delta}^1_i(s)}{ E[T_i^0]}\\
\delta^1_i(s) &=& \frac{1 - \mathbb{F}_{T_i^1}(s)}{s^2 . E[T_i^1]} + \frac{\mathbb{F}_{T_i^1}(s) \tilde{\delta}^0_i(s)}{E[T_i^1]}\\
\tilde{\delta}^0_i(s) &=& f_{T_i^0}(s) \tilde{\delta}^1_i(s)\\
\tilde{\delta}^1_i(s) &=& \frac{1 - f_{T_i^1}(s)}{s^2 } + \frac{f_{T_i^1}(s) \tilde{\delta}^0_i(s)}{E[T_i^1]}
\end{eqnarray}
\noindent where $\mathbb{F}_{T_i}(t) = 1-F_{T_i}(t)$.

Hence,
\begin{eqnarray}
 \delta^0_i(s) &=& \frac{\mathbb{F}_{T_i^0}(s)}{s^2 . E[T_i^0]}.\frac{1 - f_{T_i^1}(s) }{1- f_{T_i^1}(s) f_{T_i^0}(s)}\\
\delta^1_i(s) &=& \frac{1}{s^2 . E[T_i^1]}. \left[ 1 - \mathbb{F}_{T_i^1}(s). \frac{1- f_{T_i^0}(s) }{1- f_{T_i^1}(s) f_{T_i^0}(s)}\right]
\end{eqnarray}

And since for exponential distributions:
\begin{eqnarray*}
    f_{T_i}(s) &=& \frac{\lambda_{T_i}}{s+\lambda_{T_i}} \\
   \mathbb{F}_{T_i}(t) &=& e^{-\lambda_{T_i} t}\\
   \mathbb{F}_{T_i}(s) &=& \frac{1}{s+\lambda_{T_i}}
\end{eqnarray*}

By substitution, we get:
\begin{eqnarray}
\nonumber \delta^0_i(s) &=& \frac{\frac{1}{s+\lambda_{T_i^0}}}{\frac{s^2}{\lambda_{T_i^0}}}\cdot \frac{1-\frac{\lambda_{T_i^1}}{s+\lambda_{T_i^1}}}{1-\frac{\lambda_{T_i^1}}{s+\lambda_{T_i^1}}\cdot \frac{\lambda_{T_i^0}}{s+\lambda_{T_i^0}}}\\
        &=& \frac{1}{s^2} \cdot \frac{\lambda_{T_i^0}}{s+ (\lambda_{T_i^0}+\lambda_{T_i^1})}\\
\nonumber \delta^0_i(t) &=& \int_{y=0}^t \left[\int_{x=0}^y \lambda_{T_i^0} \cdot e^{-(\lambda_{T_i^0}+\lambda_{T_i^1})x} dx\right] dy\\
\nonumber  &=& \frac{\lambda_{T_i^0}}{(\lambda_{T_i^0}+\lambda_{T_i^1})} \int_{y=0}^t [1-e^{-(\lambda_{T_i^0}+\lambda_{T_i^1})y}] dy\\
  &=& (1-u_i) \cdot \left(t+ \frac{ e^{-(\lambda_{T_i^0}+\lambda_{T_i^1})t} - 1}{ (\lambda_{T_i^0}+\lambda_{T_i^1}) }\right)
\end{eqnarray}

And similarly,
\begin{equation}
  \delta^1_i(t) =  t-u_i \cdot \left(t+ \frac{ e^{-(\lambda_{T_i^0}+\lambda_{T_i^1})t} - 1}{ (\lambda_{T_i^0}+\lambda_{T_i^1}) }\right)
\end{equation}


\begin{thebibliography}{10}

\bibitem{Haykin}
S.~Haykin, ``Cognitive radio: brain-empowered wireless communications," \emph{IEEE JSAC},
vol.~23, no.~2, pp.~201-220, February~2005.

\bibitem{HangSu}
H.~Su and X.~Zhang, ``Opportunistic MAC Protocols for Cognitive Radio," \emph{Proc. 41st Conference on Information
Sciences and Systems (CISS 2007)}, March~2007

\bibitem{HyoilKim}
H.~Kim and K.~Shin, ``Efficient Discovery of Spectrum Opportunities with MAC-Layer Sensing in Cognitive Radio Networks," \emph{IEEE Transactions on Mobile Computing},
vol.~7, no.~5, pp.~533-545, May~2008.

\bibitem{Capacity}
S.~Srinivasa, S.~Jafar and N.~Jindal, ``On the Capacity of the Cognitive Tracking Channel," \emph{IEEE International Symposium on Information Theory}, July~2006.

\bibitem{Zhao}
Q.~Zhao, L.~Tong, A.~Swami, and Y.~Chen, ``Decentralized Cognitive MAC for Opportunistic Spectrum Access in Ad Hoc Networks: A POMDP Framework," \emph{IEEE JSAC},
vol.~25, no.~3, pp.~589-600, April~2007.



\bibitem{ZhaoWhitle}
K. Liu and Q. Zhao, ``Indexability of Restless Bandit Problems and Optimality of Whittle's Index for Dynamic Multichannel Access"
\emph{Submitted to IEEE Transactions on Information Theory},
November 2008.

\bibitem{Lifeng}
L.~Lai, H.~El-Gamal, H.~Jiang and H.~Poor, ``Cognitive Medium Access: Exploration, Exploitation and Competition," \emph{submitted to the IEEE Transactions on Networking}, October 2007.


\bibitem{HyoilKim2}
H. Kim and K. Shin, ``Fast Discovery of Spectrum Opportunities in Cognitive Radio Networks",
\emph{Proceedings of the 3rd IEEE Symposia on New Frontiers in Dynamic Spectrum Access Networks (IEEE DySPAN)}
pp. 1-12, October 2008.

\bibitem{OptimalSensing}
Won-Yeol Lee and Ian. F. Akyildiz, ``Optimal Spectrum Sensing Framework for Cognitive Radio Networks,"
\emph{IEEE Transactions on Wireless Communications},
vol. 7, no. 10, October 2008.

\bibitem{SensingTime}
Zhi Quan, Shuguang Cui, H. Vincent Poor and H. Sayed, ``Collaborative wideband sensing for cognitive radios,"
\emph{IEEE, Signal Processing Magazine},
vol. 25, no. 6, pp. 60-73, November 2008.

\bibitem{Motamedi}
A. Motamedi and A. Bahai, ``Dynamic channel selection for spectrum sharing in unlicensed bands,"
\emph{European Transactions on Telecommunications and Related Technologies},
 2007.

\bibitem{newZhao}
Q. Zhao and J. Ye, ``When to Quit for a New Job: Quickest Detection of Spectrum Opportunities in Multiple Channels,"
\emph{in Proc. of IEEE Military Communication Conference (MILCOM)},
 November, 2008.

\bibitem{Blind}
O.Mehanna, A.Sultan and H. El Gamal, ``Blind Cognitive MAC Protocols,"
\emph{in Proc. of the IEEE  International Conference on Communications (ICC'09)}, June 2009.

\bibitem{Ding}
S. Huang, X. Liu, and Z. Ding,``Optimal Sensing-Transmission Structure for Dynamic Spectrum Access,"
\emph{in IEEE INFOCOM '09}, April 2009.

\bibitem{Soft}
S. Srinivasa and S.A Jafar, ``Soft Sensing and Optimal Power Control for Cognitive Radio," in Proc. IEEE Global Commun. Conf. (Globecom), Dec. 2007.

\bibitem{Cox}
D.R. Cox, ``Renewal Theory,"
 Butler and Tanner, 1967.

\bibitem{Guha}
S. Guha and K. Munagala, ``Approximation algorithms for partial-information based stochastic control with Markovian
rewards,"
\emph{in Proc. 48th IEEE Symposium on Foundations of Computer Science (FOCS)},
2007.

\end{thebibliography}
\end{document}